\documentclass[preprint,12pt]{elsarticle}
\usepackage{dcolumn}%
\usepackage{bm}%
\usepackage{graphicx}
\usepackage{amsfonts}
\usepackage{amsmath}
\usepackage{amssymb}
\usepackage{hyperref}
\usepackage{indentfirst}
\usepackage[usenames,dvipsnames]{pstricks}
\usepackage{epsfig}
\usepackage{pst-grad} 
\usepackage{pst-plot} 
\journal{Physics Letters A}

\begin{document}
\begin{frontmatter}

\title{Topological magnetic solitons on a paraboloidal shell}
\author[1]{Priscila S. C. Vilas-Boas}
\address[1]{Universidade do Estado da Bahia, Campus VII, \\BR 402, 48970-000, Senhor do Bonfim, BA, Brasil}
\author[2]{Ricardo G. Elias}
\author[2]{Dora Altbir}
\address[2]{Departamento de F\'isica, Universidad de Santiago de Chile and CEDENNA,\\ Avda. Ecuador 3493, Santiago, Chile}
\author[3]{\\Jakson M. Fonseca}
\address[3]{Universidade Federal de Vi\c cosa, Departamento de F\'isica, 
Avenida Peter Henry Rolfs s/n, 36570-000, Vi\c cosa, MG, Brasil}
\author[2,4]{Vagson L. Carvalho-Santos}
\address[4]{Instituto Federal de Educa\c c\~ao, Ci\^encia e Tecnologia Baiano - Campus 
Senhor do Bonfim, \\Km 04 Estrada da Igara, 48970-000 Senhor do Bonfim, Bahia, Brazil}

\begin{abstract}

We study the influence of curvature on the exchange energy of skyrmions and vortices on a paraboloidal surface. It is shown that such structures appear as excitations of the Heisenberg model, presenting topological stability, unlike what happens on other simply-connected geometries such as pseudospheres. We also show that the skyrmion width depends on the geometrical parameters of the paraboloid. The presence of a magnetic field leads to the appearance of $2\pi$-skyrmions, introducing a new characteristic length into the system. Regarding vortices, the geometrical parameters of the paraboloid play an important role in the exchange energy of this excitation.

%

\end{abstract}
\end{frontmatter}

\section{Introduction}

Vortices and skyrmions are topological structures that appear in several condensed matter systems like  nematic liquid crystals \cite{Pu-Op-2013}, Bose-Einstein condensates \cite{Skyrmion-Bose-Einstein} and superconductors \cite{Mermin-Review}, among others. In magnetic systems, skyrmions appear as collective modes of the spins, forming a chiral structure with a whirling configuration. These particle-like excitations are called topological because their structures cannot be continuously deformed to ferromagnetic or another magnetic state. Several experimental works have reported the appearance and control of skyrmions in chiral magnets \cite{Experiment-skyrmions}. From the theoretical point of view, it has been shown that skyrmions can be created by constrictions in helical magnets \cite{Nagaosa-Nature-2013} and their width and chirality can be controlled by spin polarized currents \cite{Shibata-Nature-2013}. Magnetic vortices consist of a spin closure texture that can appear as the magnetization groundstate of circular nanomagnets \cite{Cowburn-Work, Hertel}. The control of the polarity and chirality of vortices as well as their pinning mechanisms have been intensively studied (See for example Refs. \cite{Vortices-Works}).

Curvature plays an important role in the properties of condensed matter systems at the nanometer scale. It is worth noting that surface shape determines the physical properties of several systems, such as the ordering of nematic liquid crystals \cite{Napoli-PRL-2012}, the local density of states in planar graphene and in defects such as disclinations or Stone-Wales defects \cite{jakson-graphene,Cortijo-NPB-2007}. Furthermore,  topological two-dimensional vortices are repelled by regions of positive Gaussian curvature and attracted by regions with negative ones \cite{Vitelli-PRL-2004}. Similarly, skyrmion width changes are induced even when the surface presents two or more characteristic length scales \cite{Dandoloff-JPA-2011,Dandoloff-PRL-1995}.

In recent years it has been shown that ferromagnetic toroidal nanorings can support vortex-like magnetization at smaller sizes than cylindrical nanorings \cite{Vagson-JAP-2010}. Furthermore, it is known that curved defects cause chiral symmetry breaking in the gyrotropic motion of vortices \cite{Van-NJP-2009}. The breaking may be originated on the interactions between the vortex core and such curved defects, leading to the generation of pinning mechanisms. Chen \textit{et al.} \cite{Chen-PRL-2012} showed that curved defects can produce changes in the vortex core diameter, determining the minimum pinning energy of vortex motion. Several models have been proposed to study interactions between magnetic vortices and curvature, considering them as Dirac's delta function \cite{Apolonio-JPA-2009}, holes \cite{Ricardo-PRB-2008}, cones \cite{cone} and hemispherical shells \cite{Gaididei-Spin-2013}. While these models are an important contribution to the subject, they do not adequately
  consider the geometry of the defects.

More recently, curved nanostructures have been constructed. Some examples are permalloy caps on non-magnetic spheres \cite{Streubel-APL-2012}, cylindrically curved permalloy magnetic segments with different radii of curvature on non-magnetic rolled-up membranes \cite{Streubel-Nanoletters-2012}, and the synthesis of periodically modulated nanowires \cite{knielsch-works}. Although paraboloidal magnetic nanostructures have not yet been developed, it has been shown that paraboloidal crystals can be constructed by an assemblage of air bubbles floating on the surface of a soap-water solution in a rotating cylindrical vessel \cite{Bragg-PSL-1947}. More recently, computer assisted design models have been used to construct ellipsoidal shapes, with thicknesses on the order of 500 nm \cite{Lazarus-PRL-2012}. Thus, by using atomic layer deposition techniques, we can expect that paraboloidally-shaped magnetic structures can soon be produced. Regarding recent theoretical works, the study of defects on paraboloidal manifolds has shown that the presence of a variable Gaussian curvature and the additional constraint of a boundary give rise to a rich variety of phenomena not found on surfaces with constant Gaussian curvature and without boundaries \cite{Giomi-PRB-2007}. 

Based on the above, in this paper we study spin excitations appearing from the exchange energy of a magnetic parabolic structure described by the Heisenberg model on a curved surface. Although we do not include magnetostatic energies in our calculations, the consideration  of  exchange  terms is  an usual study that sheds light on the behavior of magnetic systems \cite{Belavin-JETP-1975}. Therefore, the present work contributes to the study of  the stability of skyrmions and vortices on curved simply-connected manifolds, giving us results that can be considered as a first order contribution to describe magnetic properties of curved systems. In addition, the paraboloidal geometry can be used as a more realistic description of defects appearing during nano magnets fabrication. Our results show that the relation between the height and the maximum radius of the paraboloid has a significant effect on skyrmion width and vortex energy, yielding to the confinement of the skyrmion on the paraboloid top and vortex repulsion due to the positive curvature of this surface. Moreover, analytical and numerical solutions are offered to discuss the interaction with an external magnetic field. 

This work is organized as follows: in section \ref{sec2} we discuss exchange and Zeeman energies on curved surfaces and the application of the described model to the surface of the paraboloid. As well, we present analytical and numerical results of the derived equations. Finally, in section \ref{conclusions} we present the conclusions and prospects of this work.

\section{Exchange energy on a paraboloid}\label{sec2}

A paraboloid is a positive surface with variable Gaussian curvature  topologically equivalent to a plane. There are two kinds of paraboloids: elliptic and hyperbolic. In this work  we focus on circular paraboloids, which can be generated by the revolution of a parabola. Their shapes are similar to an oval cup  and can be described by the equation ${x^2}/{R^2}+{y^2}/{R^2}=z$. Although physically realizable curved thin nanostructures have a thicknesses $L$, we derive an effective two-dimensional exchange energy of a paraboloidal shell as a limiting case of $L\rightarrow0$.

\begin{figure}
\includegraphics[scale=0.5]{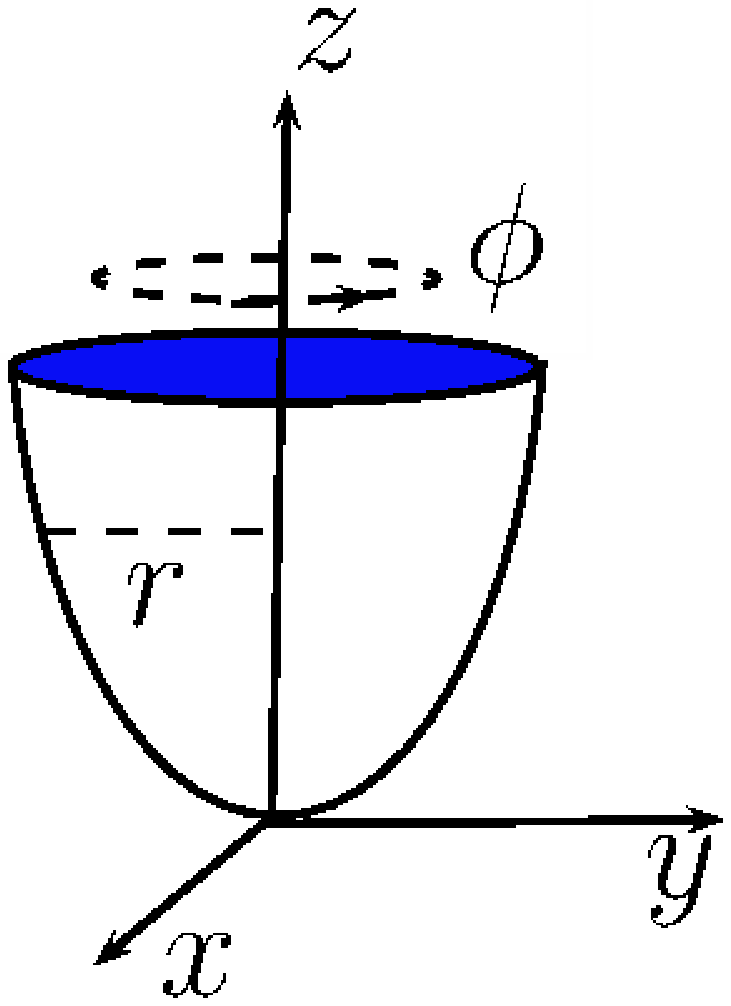}\includegraphics[scale=0.38]{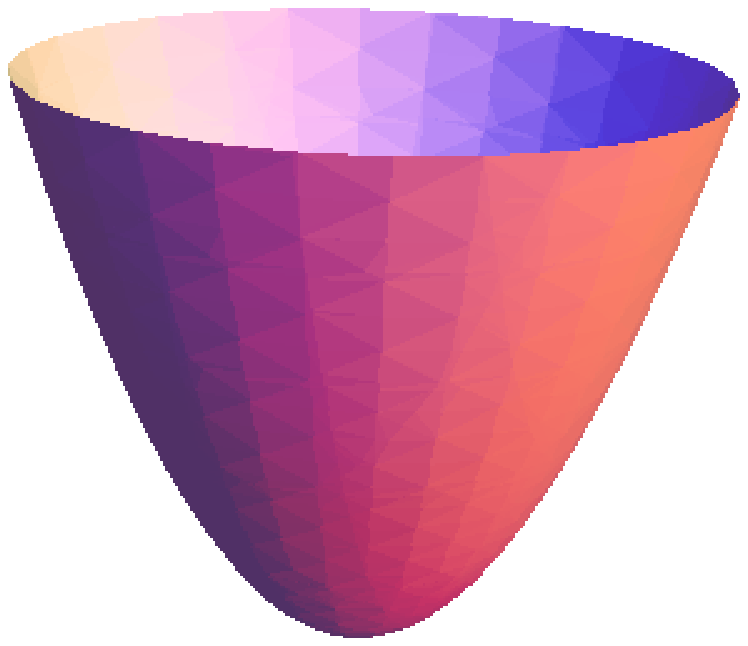}
\caption{A paraboloidal surface and the coordinate system used in this work. Note that the paraboloidal shape is invariant respect to the azimuthal angle, $\phi$. Thus, this is defined here as a cylindrically symmetric surface.}\label{ParabShape}
\end{figure}

One of the possible parametric equations representing an elliptic paraboloid is given by (See Fig. \ref{ParabShape})
\begin{eqnarray}\label{ParamParab}
x=r\cos\varphi,\hspace{0.5cm}y= r\sin\varphi,\hspace{0.5cm}z=\frac{\kappa}{2}{r^2},
\end{eqnarray}
where $h$ and $R$ are the height and maximum radius of the paraboloid, respectively, and $\kappa=2h/R^2$ is a geometric parameter that allows us to continuously deform the paraboloid into a plane by taking $\kappa\rightarrow0$. This parametrization gives the following covariant metric elements
\begin{eqnarray}\label{metrics}
g_{ij}=\left[
\begin{array}{cccc}
1+\kappa^2r^2 & 0\\
0 & r^2\\
\end{array}
\right].\end{eqnarray}

In the continuum approach of spatial and spin variables, the exchange energy of a spin system lying on a surface embedded in a 3D-space can be written as follows
\begin{eqnarray}
\label{heiscont} 
H_{_\text{EX}}=J\iint_S\left[(\nabla_{_S} m_x)^2+(\nabla_{_S} m_y)^2+(\nabla_{_S} m_z)^2\right]da\,,
\end{eqnarray} 
where $J$ denotes the coupling between neighboring spins, and according to $J<0$ or $J>0$, the Hamiltonian describes a ferro or antiferromagnetic system. The surface is described by orthogonal curvilinear coordinates $q_{1}$ and $q_{2}$. The two-dimensional gradient operator is given by
\begin{eqnarray}
\nabla_{_S}=\mathbf{e}_1\frac{1}{\sqrt{g_{11}}}\frac{\partial}{\partial q_1} +\mathbf{e}_2\frac{1}{\sqrt{g_{22}}}\frac{\partial}{\partial q_2}\,,
\end{eqnarray} 
and the surface element is $da=\sqrt{g_{11}g_{22}}dq_1dq_2$, where $g_{ii}$ is the contravariant surface metric and $\mathbf{e}_i$, $i=1,\,2$ is the unitary tangent vector. The classical spin vector field is parametrized by $\vec{m}=(m_{x},m_{y},m_{z})\equiv(\sin\Theta\cos\Phi,\sin\Theta\sin\Phi,\cos\Theta)$, so that $\Theta=\Theta(q_{1},q_{2})$ and $\Phi=\Phi(q_{1},q_{2})$. 

The interaction with an external magnetic field can be modeled by adding to the Hamiltonian (\ref{HamGen}) the term
\begin{eqnarray}\label{HamMagInt}
H_{_\text{INT}}=-g\mu\iint\mathbf{m}\cdot\mathbf{B}\,da\,,
\end{eqnarray}
where $\mathbf{B}$ is the applied magnetic field, $\mu$ is the magnetic moment, and $g$ is the Land\'e factor of the electrons in the magnetic material. 

The  Heisenberg model on a curved surface in the presence of an external field has been studied before on non-simply connected surfaces. It has been shown that tuning the magnetic field to the surface curvature in the form $B\propto1/r^2$ yields the homogeneous double sine-Gordon equation (DSGE) \cite{Vagson-Dandoloff-PLA,Vagson-Dandoloff-BJP,Vagson-Altbir-Submitted}. The associated solutions are $2\pi$-skyrmions, whose widths depend on a second length scale,  introduced into the system by the magnetic field. However, analytical results have not been obtained for other magnetic field distributions and numerical solutions must be performed \cite{Dandoloff-JPA-2011}.

The form of the magnetic field proposed in the above cited works presents a restriction to simply connected manifolds, which is associated with a divergence appearing in the magnetic field at $r=0$.  Thus, in this paper we show that this problem can be circumvented and the range of the studied model can be extended in order to be applied to simply-connected surfaces, in particular, on a paraboloid. Moreover, magnetic fields varying with $1/r^2$ are not easily produced. In this context, we perform numerical solutions for some specific and experimentally feasible fields.

By using the parametrization (\ref{ParamParab}), the total energy, given by $H=H_{_\text{EX}}+H_{_\text{INT}}$ can be rewritten as
\begin{eqnarray}\label{HamGen}
H=J\iint\Big\{\sqrt{\frac{g_{\phi\phi}}{g_{rr}}}\left[(\partial_{r}
\Theta)^2+\sin^2\Theta(\partial_{r}\Phi)^2\right]\nonumber\\
+\sqrt{\frac{g_{rr}}{g_{\phi\phi}}}\left[(\partial_{\phi}
\Theta)^2+\sin^{2}\Theta(\partial_{\phi}\Phi)^2\right]
+{\sqrt{g_{rr}g_{\phi\phi}}}\,\,\mathbf{B}'\cdot\mathbf{m}\Big\}dr d\phi\,,
\end{eqnarray}
and the derived Euler-Lagrange equations (ELE) give
\begin{eqnarray}\label{GenForm}
2(\partial^2_{\zeta}\Theta+\partial^2_{\phi}\Theta)=\sin2\Theta\left[\left(\partial_{\zeta}
\Phi\right)^2+\left(\partial_{\phi}\Phi\right)^2\right]+g_{\phi\phi}\mathbf{B}'\cdot
\partial_\Theta\mathbf{m}\,
\end{eqnarray} 
and
\begin{eqnarray}\label{phieq}
\sin^2\Theta\left(\partial^{2}_{\zeta}\Phi+\partial^{2}_{\phi}\Phi\right)+\sin2\Theta\left(
\partial_{\zeta}\Theta\partial_\zeta\Phi+\partial_{\phi}\Theta\partial_\phi\Phi\right)
=g_{\phi\phi}\mathbf{B}'\cdot\partial_\Phi\mathbf{m}\, ,\,\,\,\,\,\,
\end{eqnarray}
where $\mathbf{B}'=g\mu\mathbf{B}$. 

Due to the cylindrical symmetry $\partial_{\phi}(\sqrt{g_{\phi\phi}/g_{rr}})=0$. Furthermore, we have defined the dimensionless length scale $\zeta=\int\sqrt{{g_{rr}}/{g_{\phi\phi}}}\,dr$, which depends on the geometric parameters of the underlying manifold, explicitly written for a paraboloid as
\begin{eqnarray}\label{zetapar}
\zeta=\sqrt{1+\kappa^2r^2}-\ln\left[\frac{r_0}{r}\left(1+\sqrt{{1}+\kappa^2{r^2}}\right)\right]\,,
\end{eqnarray}
where $r_0$ is a constant of integration with units of length. It is worth noting that the above equation can give us the planar dimensionless length scale by taking $\kappa\rightarrow0$. Indeed in the case of a planar surface, $\zeta_\text{plane}=\ln\,(r/r_0)$, which can be obtained by taking $\kappa=0$ in Eq. (\ref{zetapar}) and by inserting all the remaining constants in $r_0$.

\begin{figure}
\includegraphics[scale=0.3]{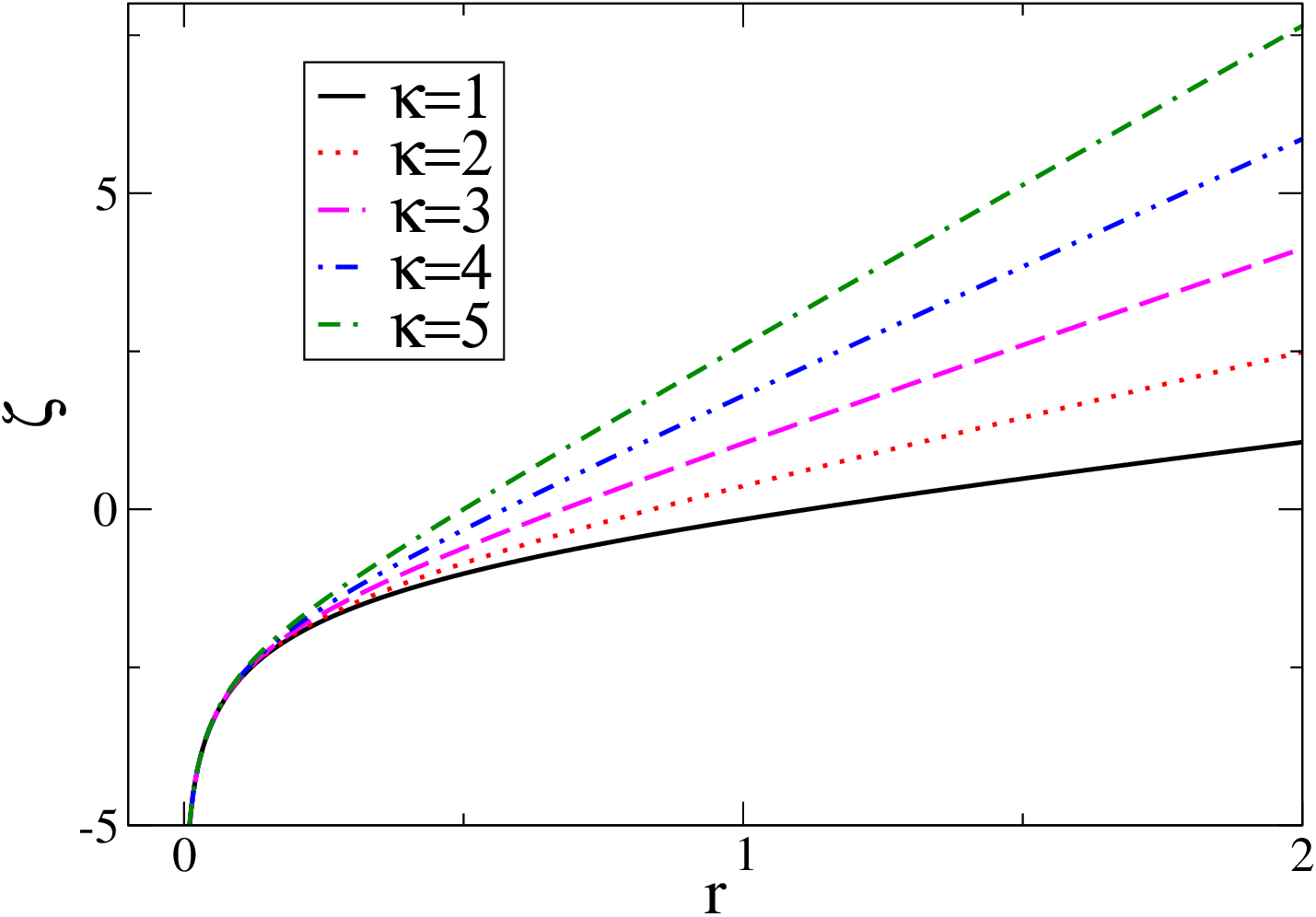}\caption{Dimensionless length scale of the paraboloidal surface, $\zeta$, for different values of $\kappa$. The interval $r\in[0,\infty)$ yields $\zeta\in(-\infty,\infty)$. $\kappa$ plays an important role in the characteristic length associated with the paraboloidal surface, such that the greater $\kappa$, the greater the value of $\zeta$.}\label{zetafunction}
\end{figure}

Fig. \ref{zetafunction} shows $\zeta$ in function of $r$ obtained from Eq. (\ref{zetapar}). In this figure, the asymptotic limits to the dimensionless length scale of the paraboloid presents the interesting property that $\zeta(r\rightarrow0)\rightarrow-\infty$ and $\zeta(r\rightarrow\infty)\rightarrow\infty$. This behavior is related to the large values of the logarithmic term for small radii. For large $r$, the first term of Eq. (\ref{zetapar}) dominates and the dimensionless length scale behaves as a linear function of $r$. The geometrical parameter plays an important role, since increasing $\kappa$ rises the $\zeta$ value as well.

\subsection{Skyrmions on the paraboloid}

In this section we discuss some aspects of skyrmions appearing from applying the described model to a paraboloidal surface. Initially we analyze the case with no external magnetic field. An analytical solution to  equations (\ref{GenForm}) and (\ref{phieq}) can be obtained by assuming cylindrical symmetry, i.e., $\Theta(r,\phi)=\Theta(r)$ and $\Phi(r,\phi)=\Phi(\phi)$. Thus, the Hamiltonian (\ref{HamGen}) can be written as
 \begin{eqnarray}\label{SimpHam}
 H=J\iint\left[(\partial_{\zeta}
\Theta)^2+\sin^{2}\Theta(\partial_{\phi}\Phi)^2\right]d\zeta d\phi\,.
\end{eqnarray}
Eqs. (\ref{GenForm}) and (\ref{phieq}) are rewritten as
\begin{eqnarray}\label{SimpTheta}
\partial^2_{\zeta}\Theta=\sin2\Theta\left(\partial_{\phi}\Phi\right)^2
\end{eqnarray}
and
\begin{eqnarray}\label{SimpPhi}
\sin^2\Theta\left(\partial^{2}_{\phi}\Phi\right)=0\,.
\end{eqnarray}

The above set of equations are identified as the sine-Gordon system (SGS), whose simplest solution consists of a skyrmion given by $\Phi=\phi+\phi_0$ and
\begin{eqnarray}\label{PiSkyrmion}
\Theta_{_\text{S}}=2\arctan\left(\text{e}^{\zeta}\right)\,,
\end{eqnarray}
The above set of equations predicts the appearance of a $\pi$-skyrmion on the paraboloid  connecting the two minima $\Theta(r\rightarrow0)=0$ and $\Theta(r\rightarrow\infty)=\pi$. 

It is also worthy to note that $\zeta$ does not determine the characteristic length of the skyrmion on the paraboloidal surface. It is a parameter valid for any surface with cylindrical symmetry that appears from the redefinition of the variables. If a skyrmion profile is given by a function $f(\lambda\zeta)$, its width is given by $\lambda^{-1}$, which generally appears in front of the $\sin2\Theta$ term of Eq. ({\ref{GenForm}). However, due to the chosen parametrization of the surface, the skyrmion width is rescaled to unity. Nevertheless, the curvature dependence of the skyrmion width can be evidenced by calculating $\zeta=\int{\sqrt{{g_{\rho\rho}}/{g_{\phi\phi}}}d\rho}$. Particularly to the paraboloid, the skyrmion width can be obtained from Eq. (\ref{zetapar}), being evaluated as $\lambda^{-1}=(r_0^{-1}+\kappa)$. The skyrmion characteristic length on a planar surface can then be calculated in the limit $\kappa\rightarrow0$. Unlike the punctured plane in which the charact
 eristic length of the skyrmion is given by the hole radius \cite{Saxena-PRB-2002}, $r_0$ is not a geometrical parameter. Indeed it is a constant of integration in such way that the skyrmion can be shrunk to a point by a continuous scaling in $r_0$, showing that skyrmion configurations are metastable in $\mathcal{R}^2$ due to the scale invariant total energy \cite{Saxena-PRB-2002,Derrick-JMP-1964}. The reason for this metastability is that there is no intrinsic length scale in the underlying manifold, as it happens in the punctured plane.

On the other hand, the skyrmion can not be shrunk to a point when $\kappa\neq0$. In fact, the increase in $\kappa$ decreases the skyrmion width, confining this structure into a small region of the paraboloid. Usually skyrmions take the length scale of the underlying manifold. However, if there are several length scales in the problem, a geometrical frustration is induced in the system and the skyrmion assume a width depending on the geometrical parameters of the surface. This phenomenon was first discussed in the context of an elastic cylinder in which periodic solitons (skyrmions) would produce deformations on the  surface in order to minimize the magnetic energy \cite{Dandoloff-PRL-1995}. Therefore, the paraboloid curvature stabilizes the skyrmion and change its width due to the two characteristic lengths associated with the paraboloid, that is, height and radius. This is an interesting result, since it indicates the possibility of controlling skyrmion 
 width by the curvature. 

The appearance of skyrmions on a paraboloid is an interesting result when this class of excitations is unstable on a surface topologically equivalent to a paraboloid, that is, a plane \cite{Derrick-JMP-1964} or a pseudosphere \cite{pseudosphere}. The difference between the stability of the skyrmion on a paraboloid and on a pseudosphere lies in the fact that, even though the pseudosphere is an infinite manifold, its negative curvature identifies it with the Poincar\'e's disc, which is a finite surface and can support only fractional skyrmion solutions, since there is a topological obstruction on this surface \cite{pseudosphere}. Moreover, unlike a paraboloid surface, the asymptotic behavior of the characteristic length of a pseudosphere is $\zeta_{\text{ps}}(\rho=0)=0$, and only half-skyrmion solutions can appear on this surface \cite{pseudosphere}.

The exchange energy of the skyrmion can be directly calculated by substituting (\ref{PiSkyrmion}) in Eq. (\ref{SimpHam}), giving
\begin{eqnarray}
E_{_\text{S}}=2\pi J\int_{\zeta{_1}}^{\zeta{_2}}[(\partial_\zeta\Theta)^2+\sin^2\Theta]d\zeta=\left.-\frac{8\pi J}{e^{2\zeta}+1}\right|_{\zeta_1}^{\zeta_2}.
\end{eqnarray}
The evaluation of the above equation in the interval $r\in[0,\infty)$ gives  $E_{_\mathcal{S}}=8\pi J$, in agreement with the Bogomolnyi inequality \cite{Bogomolnyi}.

\begin{figure}
{\includegraphics[scale=0.30]{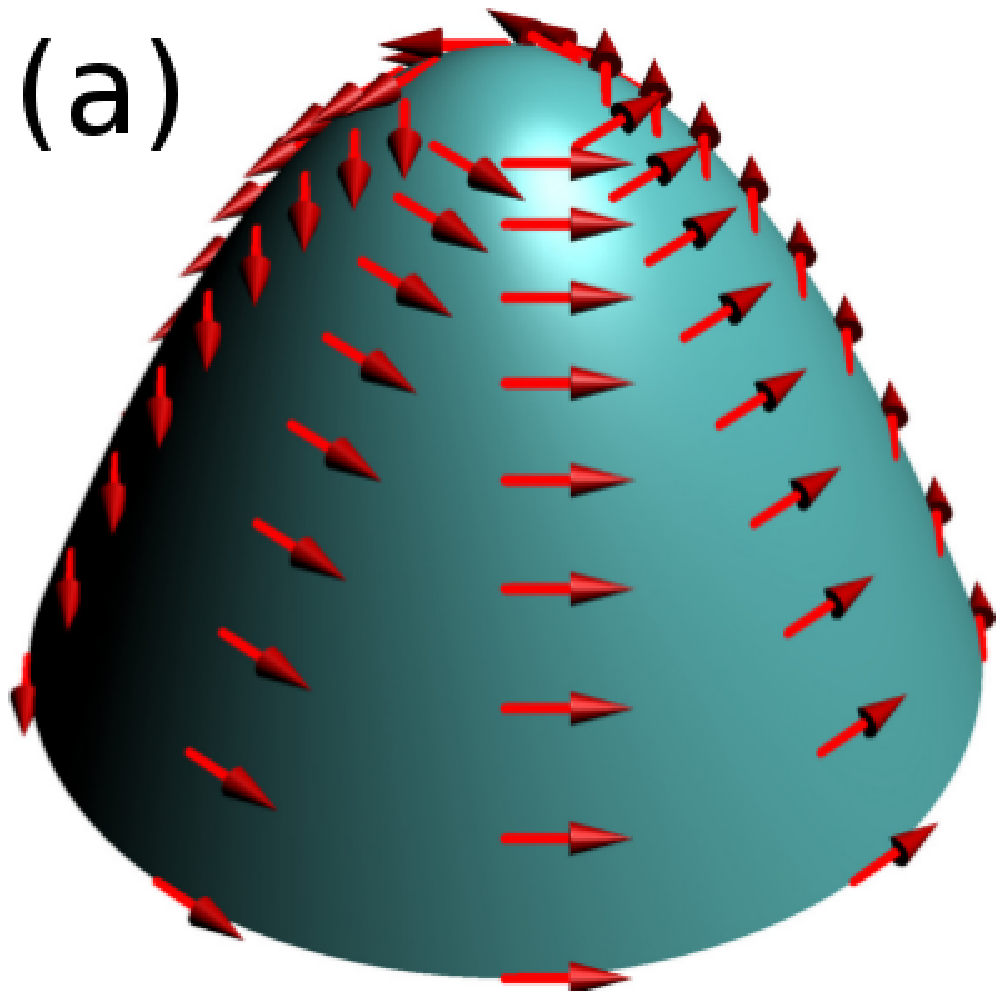}\hspace{1em}\includegraphics[scale=0.30]{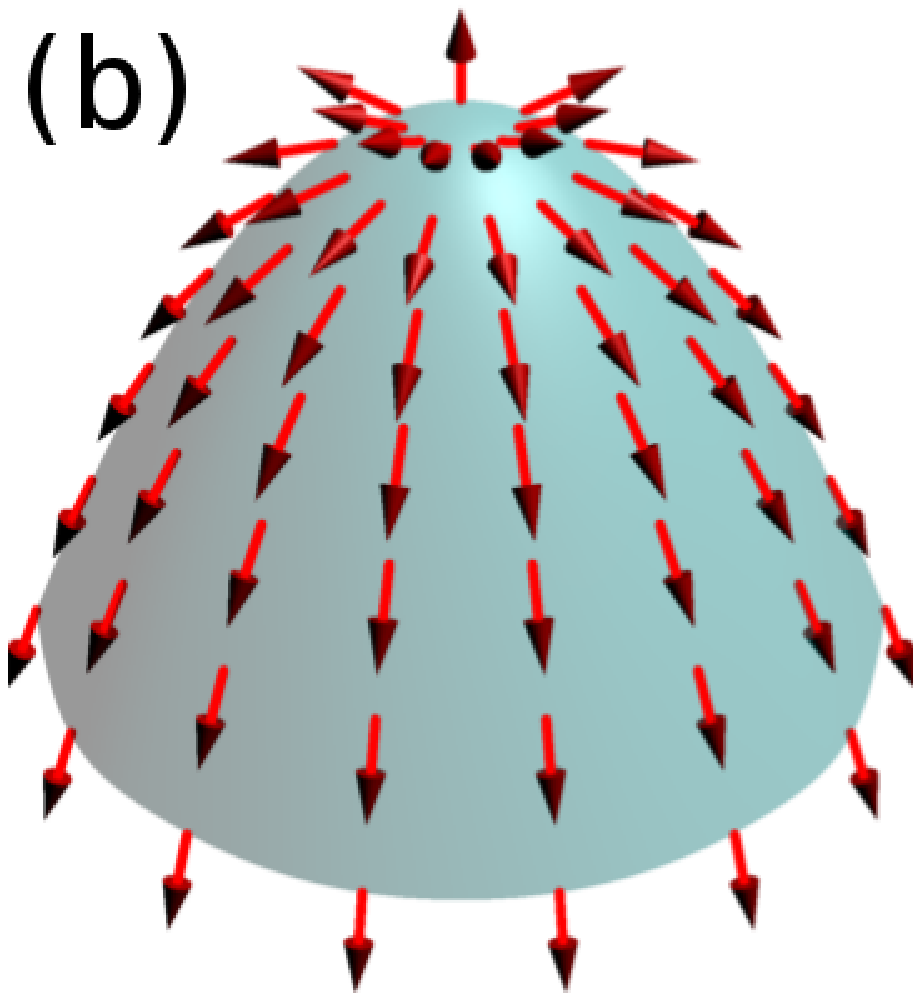}\hspace{1em}\includegraphics[scale=0.30]{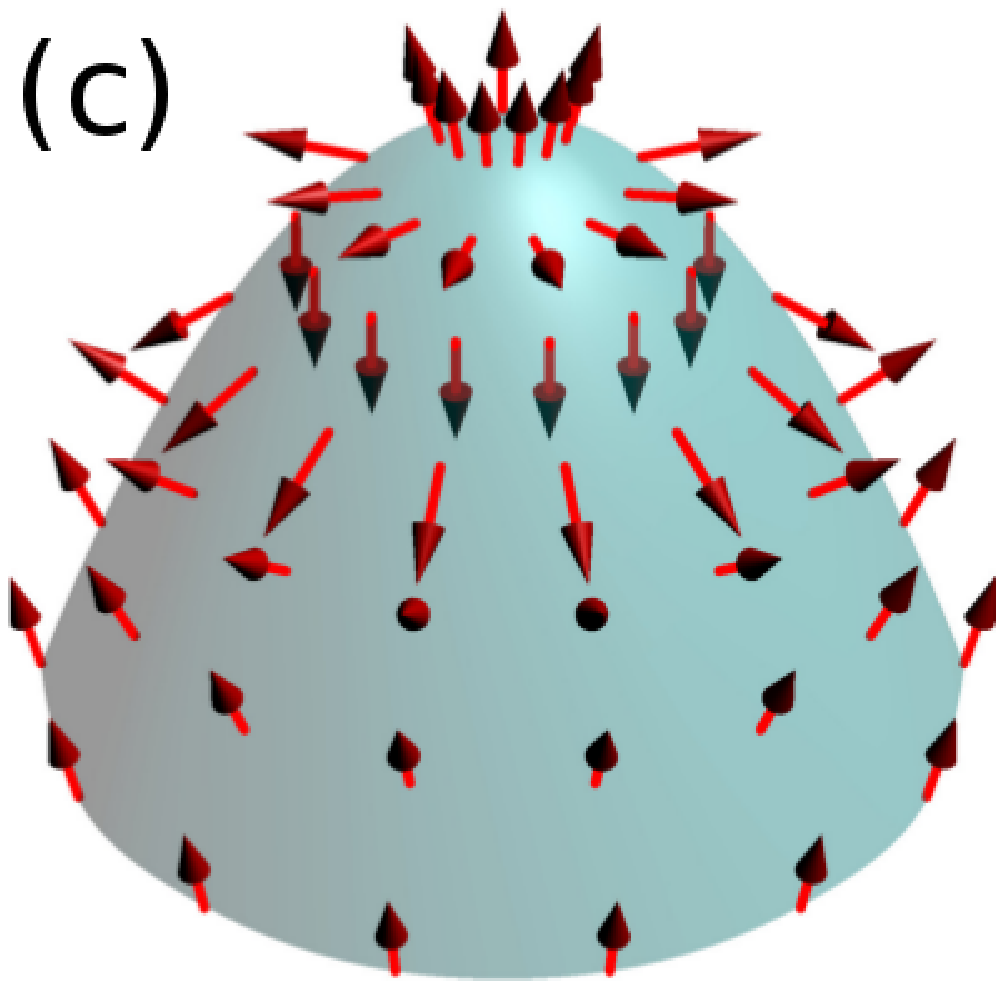}\hspace{1em}\includegraphics[scale=0.30]{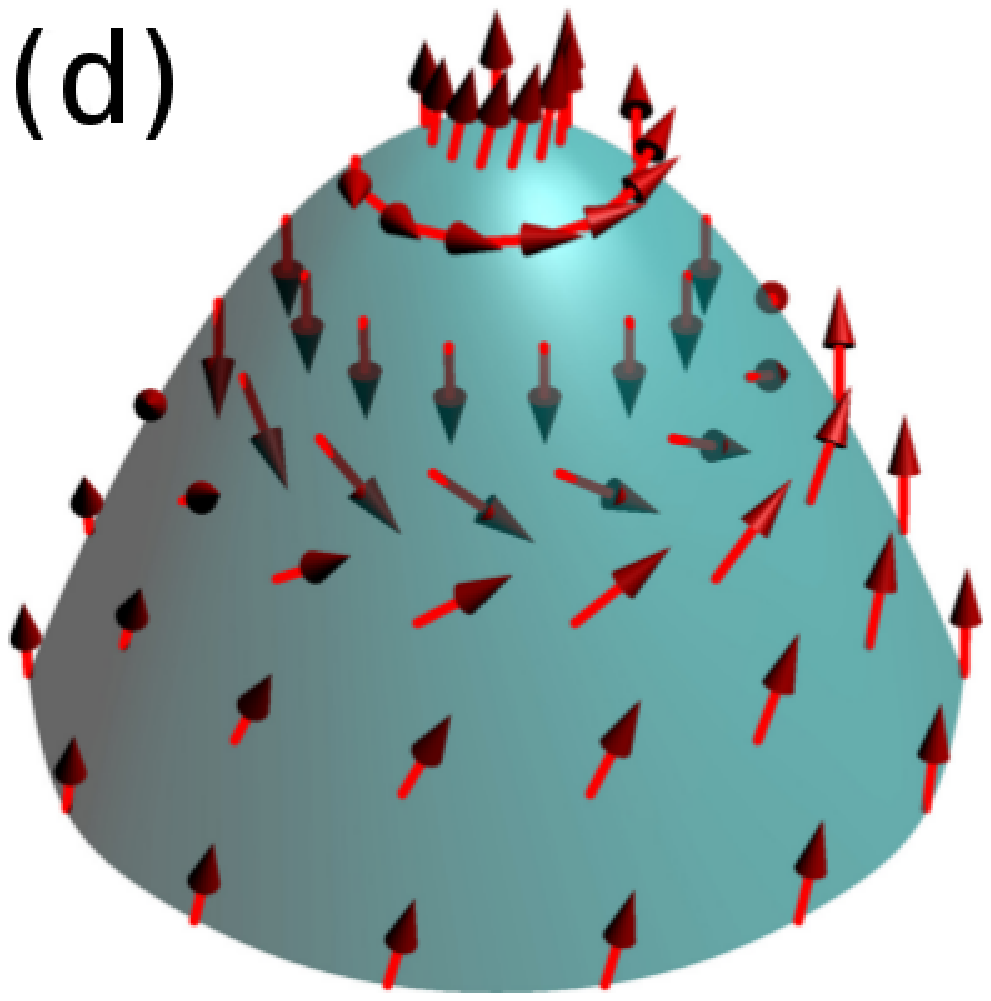}}\caption{[Color online] Possible configurations of the spin vector field on a paraboloid. Figure (a) represents a in-plane vortex with winding number $Q=1$ and phase $\phi_0=\pi/2$. Figure (b) shows a $\pi$-skyrmion described by Eq. (\ref{PiSkyrmion}), with phase $\phi_0=0$. In  figures (c) and (d), we show the two profiles for the $2\pi$-skyrmion, generated from the application of a radial and azimuthal magnetic fields, respectively.}\label{SpinPattern}
\end{figure}

If we insert a hole at the origin of the paraboloid, the winding number defined as $Q_{_\mathcal{S}}={(4\pi)^{-1}}\int\sin\Theta d\Theta d\Phi$ is given by $Q_{_\mathcal{S}}={1}/{(e^{2\zeta}+1)}$, obtaining  fractional skyrmions,  as well as other truncated non-simply-connected surfaces \cite{Saxena-PRB-2002,torusmeu,Carvalho-Santos-PLA} these solutions do not have topological stability, once the spin sphere mapping is not completely done. Similar arguments are applied when we study the above model on a paraboloid limited by its maximum radius, $r\in[0,R]$.

\subsection{Interaction with a magnetic field}

We now consider a paraboloid in the presence of a magnetic field. In previous works it was shown that a spin system described by the Heisenberg model in the presence of external magnetic fields yields a double sine-Gordon equation or the double sine-cosine-Gordon equation defined in Ref. [29], provided the magnetic field is tuned to the surface curvature in the form $B(r)=B'_{0}/r^2$, with $B'_0\equiv g\mu B_0$ constant.\cite{Vagson-Dandoloff-PLA,Vagson-Dandoloff-BJP,Vagson-Altbir-Submitted} However, the studied cases refer to non-simply connected manifolds and no considerations have been made about simply-connected surfaces for which the proposed magnetic field presents a divergence at the origin. In this context we present some arguments that ensure that the solutions discussed above are also valid on simply-connected surfaces, in particular, on a paraboloid.  

The first proposition is to introduce a cut-off of radius $\ell$ at the center of the paraboloid, which removes the divergence of the magnetic field at the origin, once it is  not necessary to define the field at this point. With the introduction of the cut-off, the punctured paraboloid becomes a non-simply-connected surface and the solutions discussed in Refs. \cite{Vagson-Dandoloff-PLA,Vagson-Dandoloff-BJP,Vagson-Altbir-Submitted} are recovered. In this case, Eqs. (\ref{GenForm}) and (\ref{phieq}) are simplified to
\begin{eqnarray}\label{ThetaEqMagExtSimp}
\partial_{\zeta}^2\Theta+\partial_{\phi}^2\Theta=\frac{\sin2\Theta}{2}\left[\left(\partial_\phi\Phi\right)^2+\left(\partial_\zeta\Phi\right)^2\right]-\frac{r^2}{2}F(r)\end{eqnarray}
and
\begin{eqnarray}\label{PhiRadEqMagExt}
\sin\Theta\left(\partial_{\zeta}^2\Phi+\partial_{\phi}^2\Phi\right)+2\cos\Theta\left(\partial_\zeta\Theta\partial_\zeta\Phi+\partial_\phi\Theta\partial_\phi\Phi\right)=-r^2 B(r)G(\phi)\,.
\end{eqnarray}
Here
\begin{eqnarray}
F(r)=\left\{\begin{array}{cccc}
B(r)\sin\Theta, & \text{for}\,\,\,\,\mathbf{B}=B(r)\hat{z}\\
-B(r)\cos\Theta\cos(\Phi-\phi)
, & \text{for}\,\,\,\,\mathbf{B}=B(r)\,\hat{\phi},\,\hat{r}\end{array}\right.
\end{eqnarray}
and
\begin{eqnarray}
G(\phi)=\left\{\begin{array}{cccc}
0, & \text{for}\,\,\,\,\mathbf{B}=B(r)\hat{z}\\
\sin(\Phi-\phi+\phi_0), & \text{for}\,\,\,\,\mathbf{B}=B(r)\,\hat{\phi},\,\hat{r}\,\end{array}\right.,
\end{eqnarray}
where $\phi_0=0$ for a radial field and $\phi_0=\pi/2$ for an azimuthal field. 

From Eqs. (15) and (16), it can be noted that the assumptions $\partial_\phi\Theta=0$ and $\partial_\zeta\Phi=0$ are not fulfilled when a generic magnetic field is considered. However, by adopting a magnetic $B(r)= B_0'/r^2$, these assumptions can be promptly used. In this way, Eq. (\ref{PhiRadEqMagExt}) admits the general solution $\Phi=\phi+\varphi_0$, where $\varphi_0$ is a phase depending on the magnetic field direction, such that $\varphi_0=0$ for a radial or an axial magnetic field and $\varphi_0=\pi/2$ for an azimuthal magnetic field. Thus, $2\pi$-skyrmions are obtained from Eq. (\ref{ThetaEqMagExtSimp}). The skyrmion width can not be defined from this general surface parametrization. In this case the skyrmion width is rescaled to 1, losing the dependence on the geometry and depending only on the strength of the magnetic field. Furthermore, due to the introduced cut-off, the solutions have fractional charge and do not describe stable topological excitations when the sp
 in sphere mapping has not been completely done. The system is then topologically unstable and homotopy arguments cannot ensure its stability.

In order to obtain stable skyrmions, another proposal is needed to avoid the magnetic field divergence at the origin. For $r=0$, Eq. (\ref{ThetaEqMagExtSimp}) is reduced to the SGS, giving us the freedom to choose the magnetic field at this point, without changing  the properties of the differential equation. Thus, the external magnetic field can be given by the relation
\begin{eqnarray}
B(r)=\left\{\begin{array}{cccc}
0, & \text{for}\,\,\,\,r=0\\
({1}/{r^2})B_0, & \text{for}\,\,\,\,r>0\,\end{array}\right.\,.
\end{eqnarray}
Now, since  no cut-off has been added, a simply-connected surface is obtained, with $\zeta(r=0)\rightarrow-\infty$. Then, the $2\pi$-skyrmion is topologically stable, that is, the spin excitation has Skyrme number $Q_{_\mathcal{S}}=2$. Therefore, this belongs to the second class of the second homotopy group and can not be deformed into a ferromagnetic state or a $\pi$-skymion by a continuous transformation. By explicitly writing Eq. (20) of Ref. [14] in terms of $r$, we obtained $\lambda^{-1}=\sqrt{2\kappa^2/(B'_0-2)}$. Then, the interplay between the magnetic field and the curvature of the paraboloid determines the skyrmion width. The presence of the magnetic field induces another characteristic length into the system, confining the  skyrmion in smaller regions for higher magnetic field values. The pattern associated with $2\pi$-skyrmion on a paraboloid is shown in Figs. \ref{SpinPattern}.(c) and (d). 

Since analytical solutions are possible only when the magnitude of the magnetic field varies with $1/r^2$, which is not easy to do experimentally, we perform numerical analysis of two feasible magnetic fields with current technology: a) constant  azimuthal magnetic field  and b) azimuthal magnetic field varying with $1/r$. Due to the symmetry of the considered magnetic fields, we will continue to use the cylindrical symmetry to the order parameters, which allows us to perform our numerical calculations only to $\Theta(r)$. Then we can rewrite Eq. (\ref{ThetaEqMagExtSimp}), explicitly for the paraboloidal case, as
\begin{eqnarray}\label{Doubl-sine-Par}
\frac{r^2}{1+\kappa^2r^2}\partial^2_r\Theta+\frac{r}{(1+\kappa^2r^2)^{2}}\partial_r\Theta=\frac{1}{2}\left[\sin2\Theta+F(r)\right]\,.
\end{eqnarray}

The numerical results we obtained refer only to the case of an azimuthal magnetic field. Results to radial and axial magnetic fields can be obtained as well, however, we do not present them in this work. The first case, a constant magnetic field, $B(r)=B_{0}\hat{\phi}$, can be generated by an electric current $I$ in a cylindrical conductor of radius $a$ with a cavity of radius $b$. In this case, the strength of the magnetic field, in the SI units, is given by $B=\mu_{0}Iy/2\pi(a^2-b^2)$, where $y$ is the distance between the center of the conductor and the center of the cavity \cite{Greiner-Book}. The equation obtained does not have analytical solutions for $B_0\neq0$, but the asymptotic behaviour for $r\ll1$  can be analyzed. In this case, the terms proportional to $r^2$ can be neglected and Eq. (\ref{Doubl-sine-Par}) is simplified to
\begin{eqnarray}\label{AsymptoticThetaI}
\partial_r\Theta=\frac{\sin2\Theta}{2r}\Rightarrow\Theta_{(\text{i})}(r\ll1)=\tan^{-1}\,(r)+\theta_0\,,
\end{eqnarray}
where $\theta_0$ is a phase that does not account to the energy calculations and is useful for discussing the numerical results. This function is independent of the external field. 

On the other hand, the asymptotic solution when $r\gg1$, Eq. (\ref{Doubl-sine-Par}) is evaluated as
\begin{eqnarray}
\partial_r^2\Theta=\frac{\kappa^2}{2}\sin2\Theta-\frac{\kappa^2r^2}{2}B_0\cos\Theta\,,
\end{eqnarray}
which has no analytical solutions either. However, by writing the energy density from Eq. (\ref{HamGen}) as $\mathcal{H}=(\partial_\zeta\Theta)^2+V(\Theta)$, its associated potential, $V(\Theta)=\kappa^2\sin^2\Theta-\kappa^2r^2B'_0\sin\Theta$, is minimized for $\Theta_{\text{min}}=(2n+1)\pi/2$, with $n$ being odd or even, depending on the magnetic field direction, $-\hat{\phi}$ or $+\hat{\phi}$, respectively. 

The numerical results are summarized in Fig. \ref{NumericosConst}. First, we assumed $\theta_0=0$, and performed numerical calculations for different magnetic field strengths. When the magnetic field increases, the amplitude of the oscillation diminishes and for large fields the spins align along $\pi/2$. We then tested what happens when the phase varies. 
Depending on $\theta_0$, the spins oscillate around one of the minima $\Theta_{\text{min}}=(4n+1)\pi/2$, but there is no appearance of an excitation to connect these minima. In this way, even for large field strengths, topological skyrmions do not appear when an azimuthal and constant magnetic field is applied on a paraboloid surface. The geometric parameter also influences the behavior of the order parameter, such that the larger the $\kappa$, the lower the deviations, as shown in Fig. \ref{NumericosConst}. In this way, the geometrical parameter acts as a magnetic field, forcing the spins to align them parallel to each other.

\begin{figure}
\includegraphics[scale=0.30]{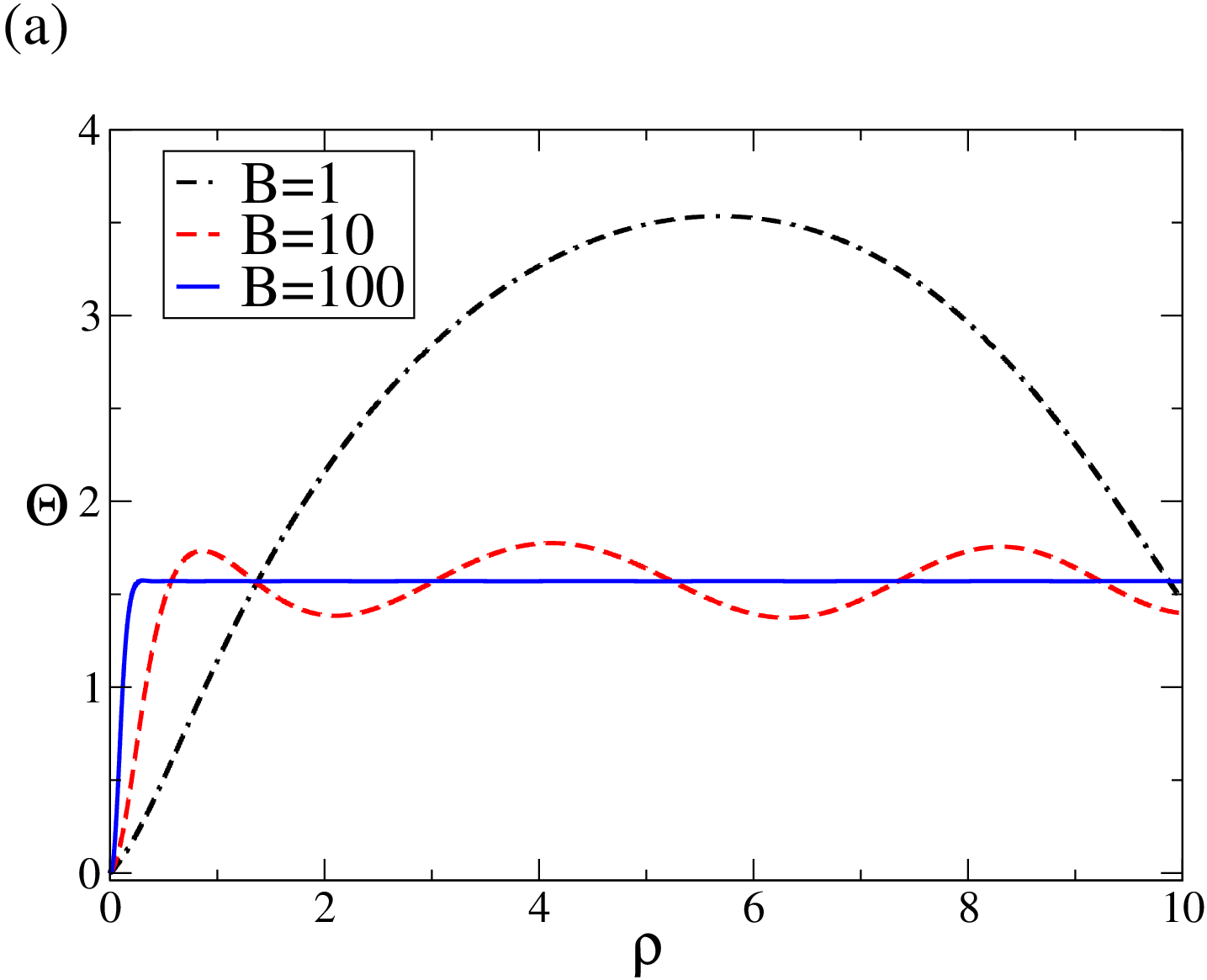}\hspace{0.2cm}\includegraphics[scale=0.30]{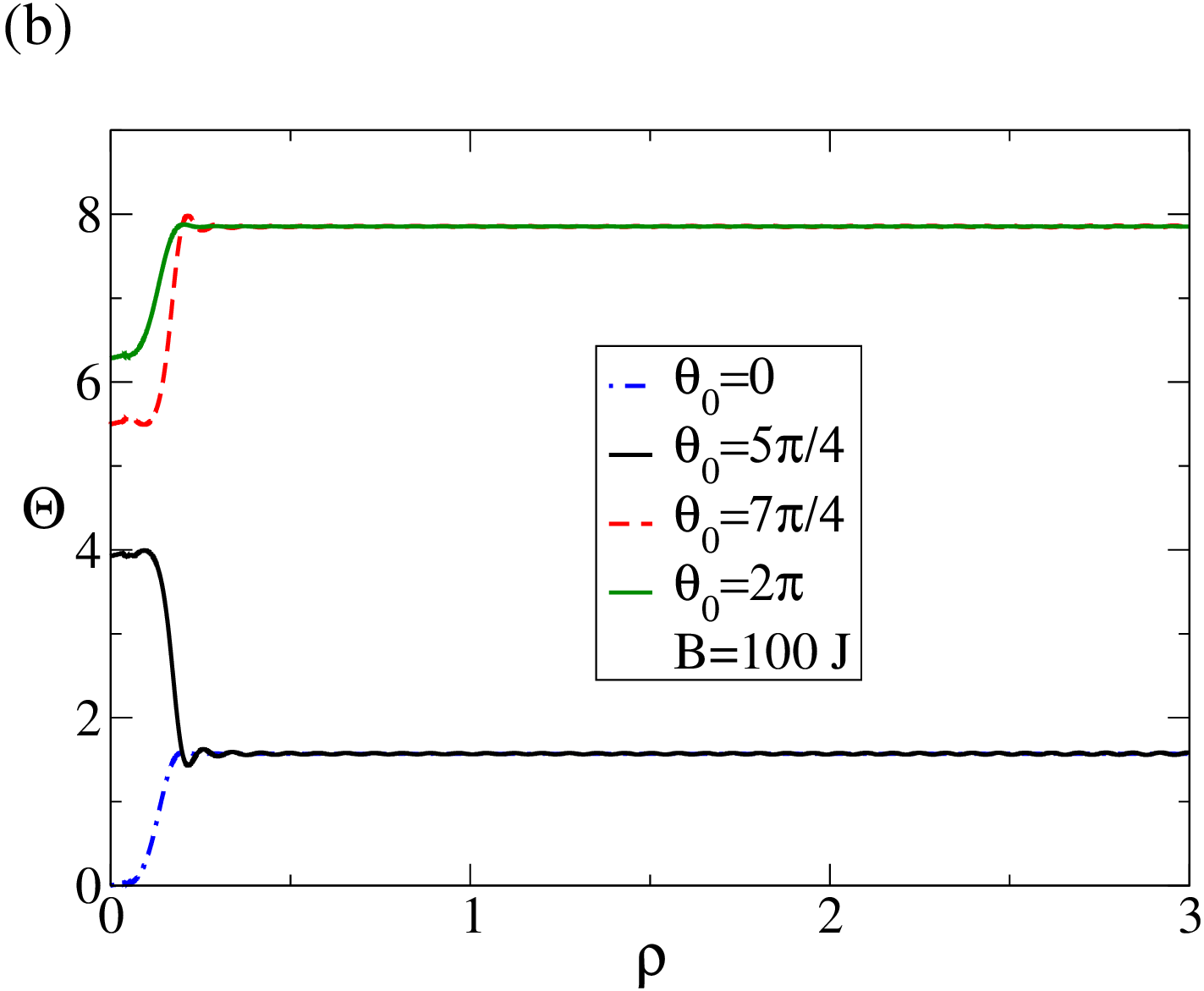}\\\\\includegraphics[scale=0.30]{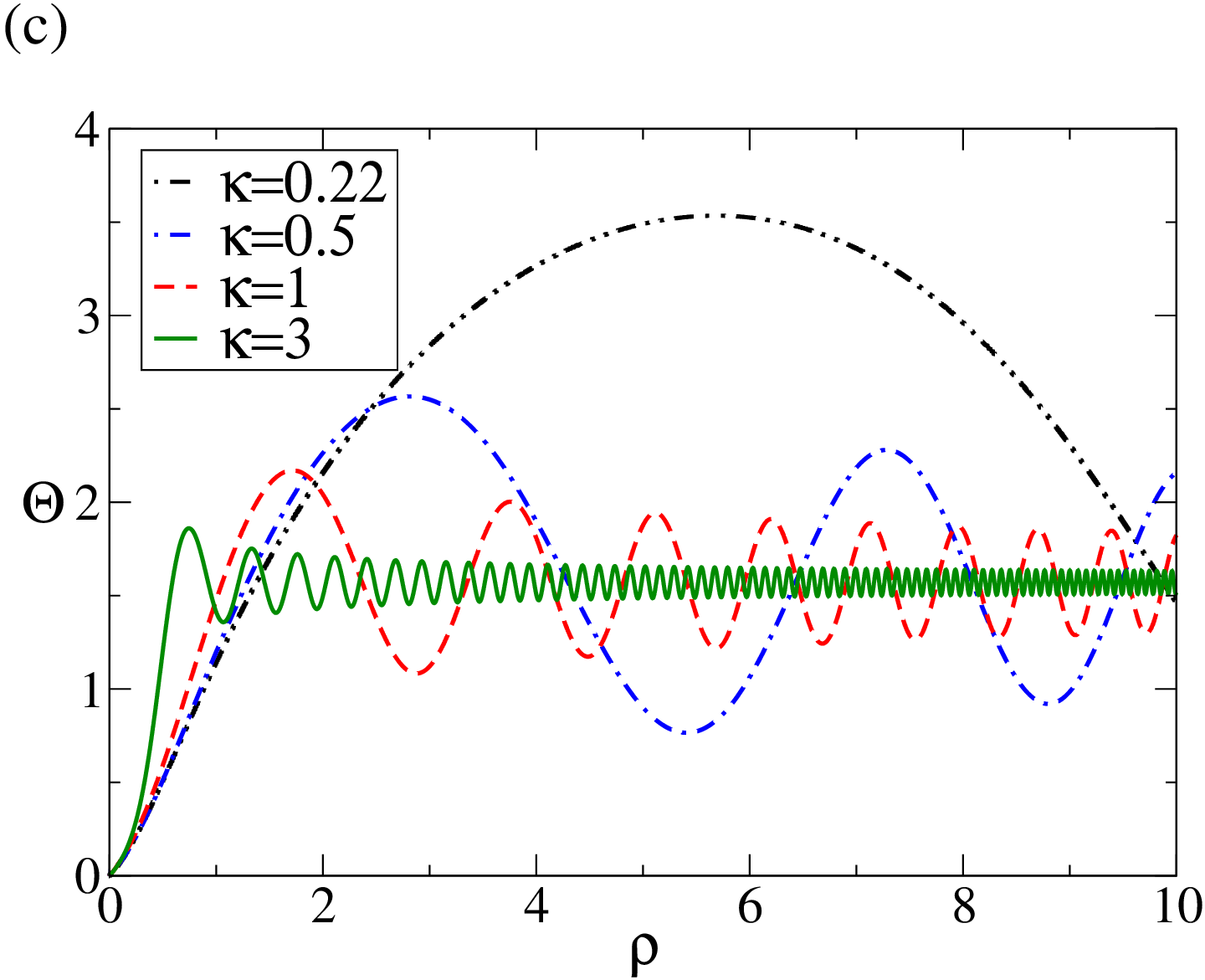}\hspace{0.2cm}\includegraphics[scale=0.30]{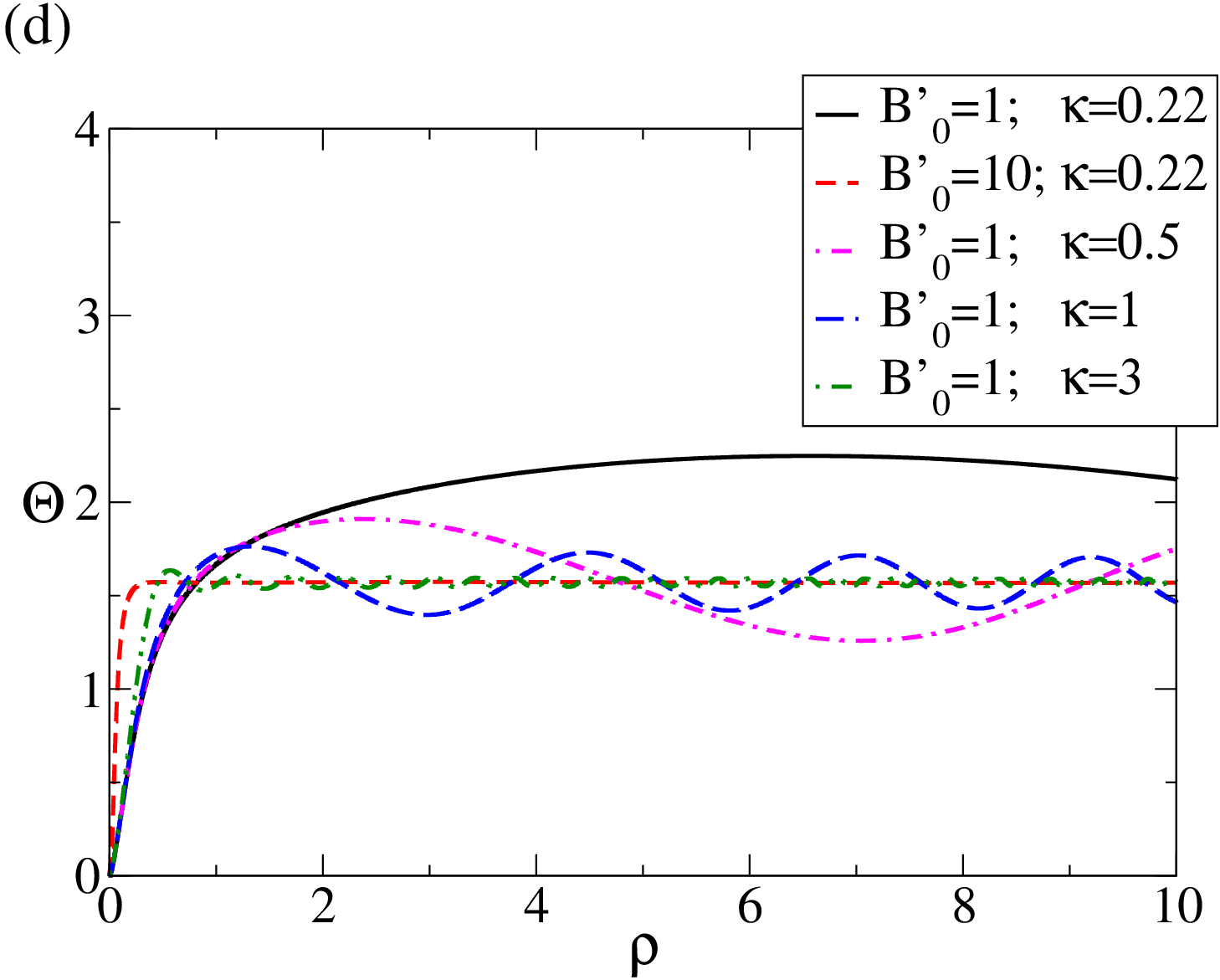}\caption{[Color online] Figures (a), (b) and (c) illustrate $\Theta$ obtained by solving numerically Eq. ({Doubl-sine-Par}) for a constant azimuthal magnetic field, with $B$  in units of $J$, $R\in[3,10]$ and $h=1$. Figure (c) depicts the behavior of the spin vector field for different geometrical parameters, $\kappa$. The geometrical parameter determines oscillation amplitudes, but does not change the qualitative behavior of the spin system. Figure (d) shows numerical results for a magnetic field varying with $1/r$.}\label{NumericosConst}
\end{figure}

Finally, we analyze the case $B = B'_0/r$.  Adopting the same procedure to find the asymptotic solutions for $r\ll1$, the simplified equation has the solution
\begin{eqnarray}\label{AsymptoticThetaII}
\Theta_{(\text{ii})}(r\ll1)=2\tan^{-1}\left[\tanh\left(-\frac{B_0r}{4}\right)\right]+\tan^{-1}\,({r})+\theta_0 \,.
\end{eqnarray}
Unlike the asymptotic solution (\ref{AsymptoticThetaI}), Eq. (\ref{AsymptoticThetaII}) depends on the magnetic field strength, which explains the behavior observed in Fig. \ref{NumericosConst}, in which a magnetic field decreasing with $r$ is more efficient to align the spins into the in-plane direction when the amplitude of the oscillations is lower.  Then, as shown in Fig. \ref{NumericosConst}, when the field varies with $1/r$, spins align along the field for a smaller field as compared to the constant magnetic field.

\subsection{Vortices on the paraboloid}
We will now return to study the case in which no magnetic field is acting into the system. In this situation another class of solutions appearing from Eq. (\ref{GenForm}) can be obtained by imposing $\Theta=\pi/2$. This condition produces as the simplest solution $\Phi=Q\phi+\phi_0$ to Eq. (\ref{SimpPhi}), consisting of a closure spin texture, also known as a vortex state. Here, $Q$ is the vortex winding number formally defined as
\begin{eqnarray}
Q=\frac{1}{2\pi}\oint_{C} (\vec{\nabla}\Phi)\cdot d\vec{l}\,.
\end{eqnarray}
This equation gives an integer number when the integration is evaluated along a closed path $C$ around the surface. However, the energy of a vortex depends on $Q^2$, such that solutions with $Q>1$ are energetically unstable \cite{Carvalho-Santos-PLA}. In this way, from now on, we consider only $Q=1$, which can be viewed in Fig. \ref{SpinPattern}-(a).

Regarding magnetic systems, the vortex pattern is a magnetization configuration arising from the competition between exchange and dipolar energies. In the present paper we analyze only the exchange energetics by considering that the dimensions of the paraboloid ensure the stability of such excitations. The energy associated with this configuration is calculated from the substitution of Eq. (\ref{zetapar}) in Eq. (\ref{HamGen}), being evaluated as
\begin{eqnarray}\label{VortexPatternEnergy}
E_{_{\text{V}}}=E_{_C}
+2\pi J\left[\Lambda+\ln\left(\frac{R(1+\sqrt{1+\kappa^2\ell^2})}{\ell(1+\sqrt{1+\kappa^2R^2})}\right)\right]\,,
\end{eqnarray}
where $E_{_C}$ is the vortex core energy, $\ell$ is the vortex core radius and $\Lambda=\sqrt{1+\kappa^2R^2}-\sqrt{1+\kappa^2\ell^2}$. The vortex core appears to prevent the divergence at the origin, in which the continuum approximation to the exchange energy fails. By adopting the rigid vortex core model \cite{New-Ref-Dora}, in which there are not changes in the size and form of the core profile due to the curvature, we can calculate the energy of this region by using the model proposed by Landeros \textit{et al} \cite{Landeros-PRB-2005}, in which $m_{_z}=[1-(r/\ell)^2]^n$. In this case, vortex core energy can be evaluated as
\begin{eqnarray}
E_{_C}=J\left(\frac{1}{2}H[2n]-nH[-\frac{1}{2n}]\right)\,,
\end{eqnarray}
where 
\begin{eqnarray}
H[z]=\sum_{i=1}^{\infty}\left(\frac{1}{i}-\frac{1}{i+z}\right)\,
\end{eqnarray}
are the harmonic numbers \cite{Landeros-PRB-2005}.

The profile of a magnetic vortex core can be only obtained from the exchange energy of a plane disk by parametrizing the magnetization vector by a complex function \cite{Metlov-JMMM-2013}. However, this out-of-plane vortex component of the magnetization vector yields a magnetostatic energy cost, coming from surface charges, formally defined as $\sigma=\vec{m}\cdot\hat{n}$. Thus, in order to analyze the stability of magnetic vortices on the paraboloid, the dipolar energy associated with different magnetization configurations and the magnetostatic energy associated to the vortex core region must be calculated. Since the exchange energy evaluated under the rigid vortex core model does not depend on the geometry of the magnetic system, our analysis focuses on the in-plane component of the vortex pattern. 

From Eq. (\ref{VortexPatternEnergy}) it can be noted that the in-plane component of the vortex energy on a paraboloid depends on $r$, evidencing a logarithmic divergence at the origin, as happens when the energy associated with this excitation is evaluated on a plane disc. Nevertheless, it also depends on the geometrical parameter of the paraboloid such that the greater the $\kappa$, the grater the vortex energy. Thus, the height of the paraboloid plays an important role in the vortex stability. Notably, lowest vortex energy is obtained for $\kappa=0$, in agreement with the result of Vitelli \textit{et al} \cite{Vitelli-PRL-2004}, in which the authors showed that a planar vortex is repelled by a positive Gaussian curvature, and attracted by a negative one. The paraboloid has positive Gaussian curvature, given by 
\begin{eqnarray}
G=\frac{\kappa^4}{4+\kappa^4r^2}\,,
\end{eqnarray}
being positive for any $\kappa\neq0$. These results are summarized in Fig. \ref{Energy}, in which we show the in-plane vortex energy on a paraboloid with external radius $R$ for different values of $\kappa$.

\begin{figure}
\includegraphics[scale=0.3]{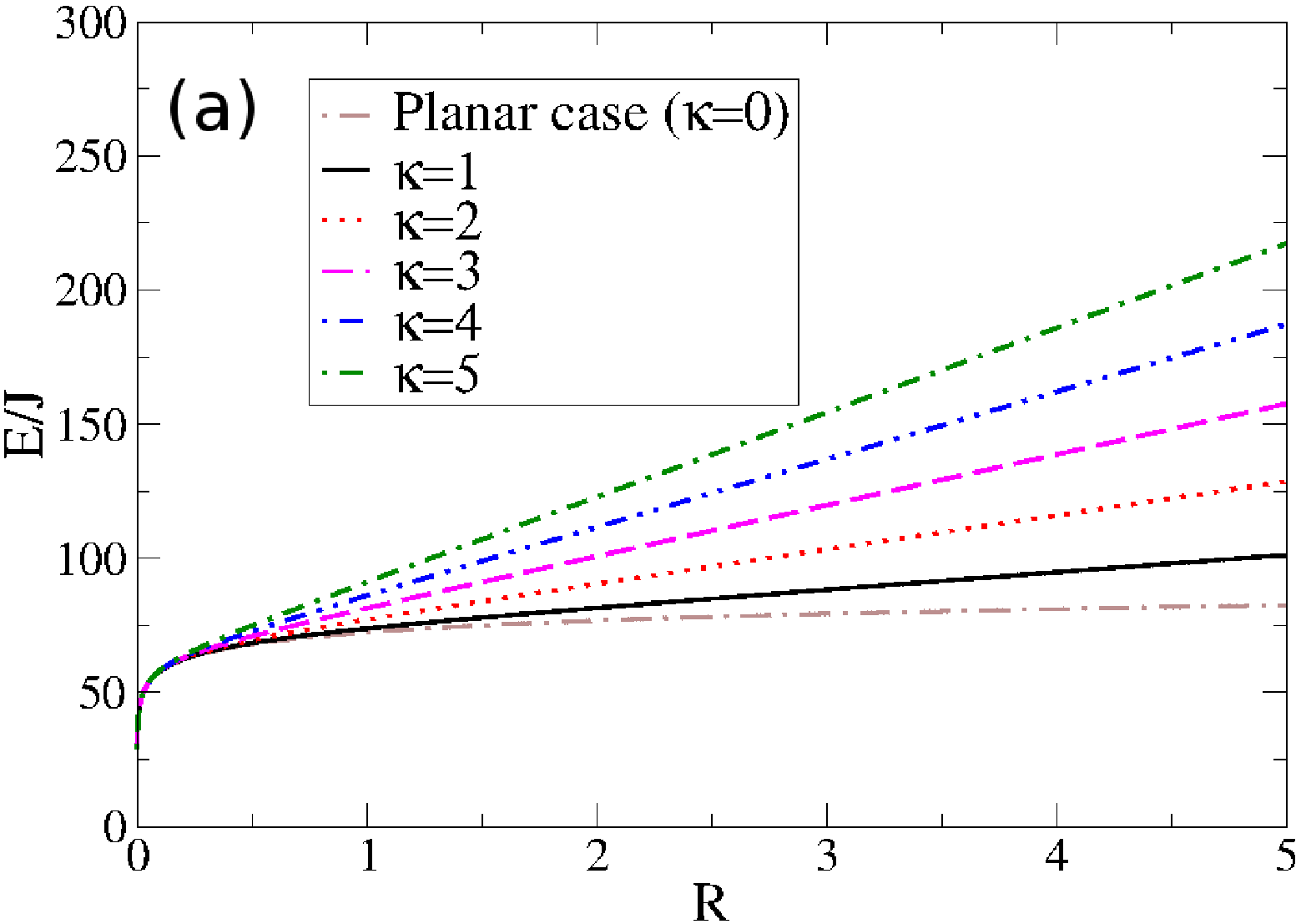}\hspace{0.5cm}\includegraphics[scale=0.3]{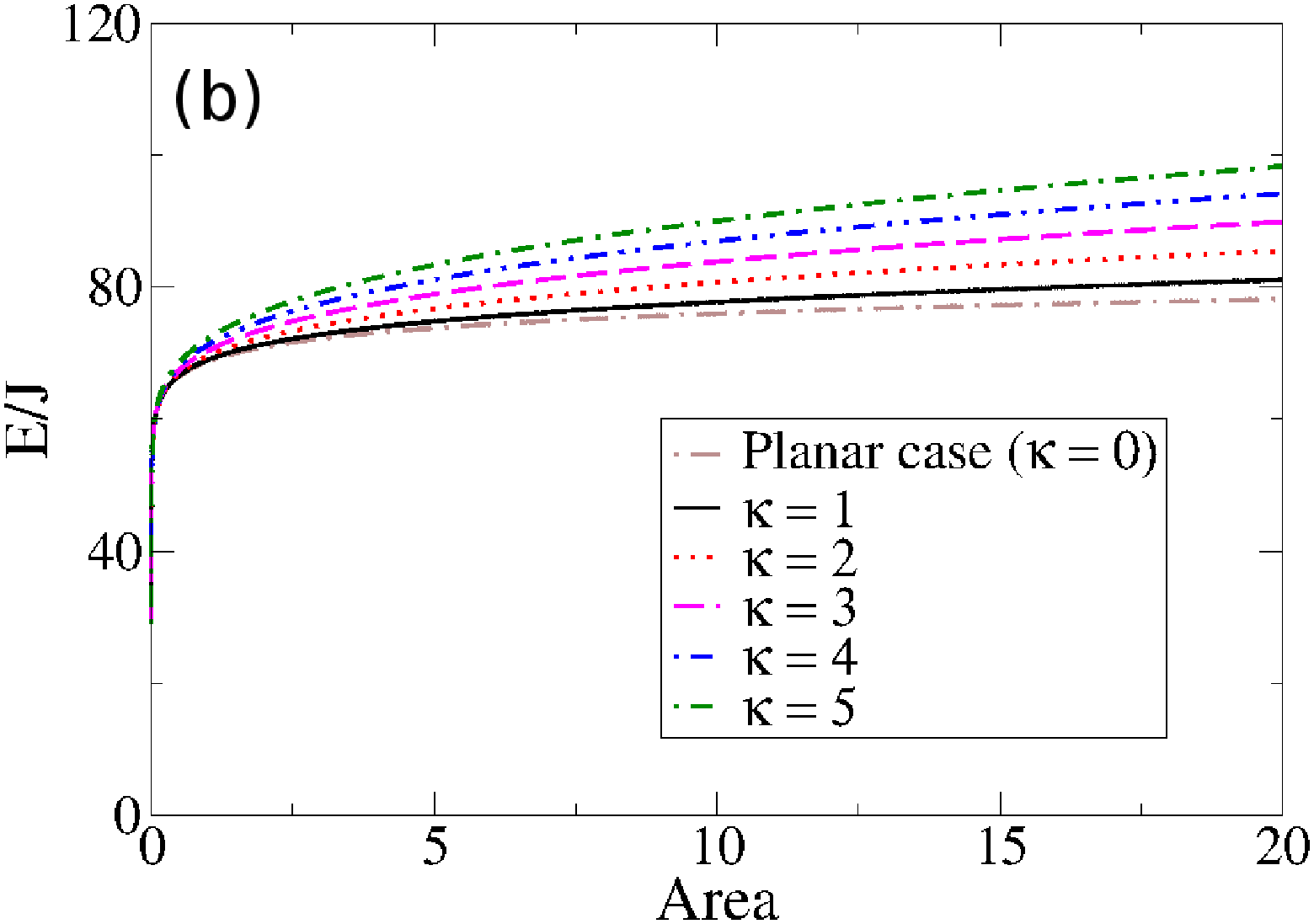}\caption{[Color online]  Energy of the vortex configuration on the paraboloid surface as a function of (a) the radius of the paraboloid, and (b) the area of the surface of the paraboloid.  In both figures $R\in[0.0001,5]$ and $\ell=0.00001$.} \label{Energy}
\end{figure}

The analysis above focuses on geometries with the same external radius, that is, the paraboloid with $\kappa\neq0$ has a larger area than a disk with the same external radius. Thus, in order to analyze the energy of the vortex state when a disk with external radius $\rho$ is deformed in order to get a paraboloid with external radius $R$ and height $h$, we calculate the relation between the external radius of a paraboloid and the disc when they have the same area. The relation between the disk radius and the paraboloidal geometrical parameter is given by
\begin{eqnarray}\label{radius}
\rho=\left\{\frac{2}{3\kappa^2}\left[\sqrt{(1+\kappa^2R^2)^3}-1\right]\right\}^{1/2}\,.
\end{eqnarray}
Figure \ref{Energy} shows the results for the vortex energy as a function of the surface area on a plane disc and on a paraboloid with different geometrical parameters. The vortex energy is always greater on the paraboloid than on its counterpart disk.

\section{Conclusions and prospects}\label{conclusions}
We studied the stability and energy of topological excitations from exchange energy on a paraboloidal surface. In the absence of an external magnetic field, the solutions to the equations of motion are $\pi$-skyrmions, whose width depends on the the geometrical parameter of the paraboloid. A change in skyrmion width induced by curvature appears in the system due to two characteristic lengths associated with the paraboloidal surface. Thus, there is the possibility of controlling the skyrmion width by using curvature effects. 

By introducing of an external magnetic field, we showed that $2\pi$-skyrmions are stable on this surface and that the magnetic field introduces a new characteristic length into the system, changing skyrmion width. We obtain numerical results for a constant magnetic field and a magnetic field varying with $1/r$ and show that these fields do not induce the appearance of skyrmions, but forces the spin to align along the magnetic field direction. The variable magnetic field is more efficient in aligning the spins, showing an interesting interplay between the curvature and magnetic field control of the properties of magnetic systems.

Finally, we calculate the in-plane vortex energy on a paraboloidal shell and showed that these spin configurations have lower energy on a plane surface than on its paraboloidal counterparts. Thus, if miniaturization is desired, an in-plane magnetic vortex comprising-systems must be formed by plane devices. These results show some intriguing aspects of the stability of vortices on curved surfaces and on the interaction of the planar vortex with curved defects appearing on nanomagnets. The results summarized in Fig. \ref{Energy} show that planar vortices have greater energy on curved magnets with paraboloidal shapes than on planar cylindrical nanomagnets with the same external radius. Thus, if the miniaturization of the magnetic nanoelements is desired, from the viewpoint of exchange energy, planar devices are more efficient. On the other hand, regarding curved defects from the fabrication of magnetic nanodots, ridges and valleys-like defects could be modeled by paraboloidal sh
 apes. Our results suggest that an in-plane vortex is pushed by the positively curved defects and pulled by negative ones. However, since the vortex core has diameters on the order of $20$ nm \cite{Metlov-JMMM-2013}, magnetostatic interactions start to be important in this length scale, in such a way that the competition between the exchange and magnetostatic energies determine the vortices pinning and depinning mechanisms induced by curved defects. A complete analysis of this problem is still pending.

\section*{acknowledgements}
V.L.C.S. thanks the Brazilian agency CNPq (Grant No. 229053/2013-0), for financial support. D.A. acknowledges the support of FONDECYT under project 1120356, the Milennium Science nucleus ``Basic and Applied Magnetism'' P10-161-F from MINECON, and Financiamento Basal para Centros Cient\'ificos y Tecnol\'ogicos de Excelencia, under project FB 0807. We also acknowledge AFOSR Grant. No. FA9550-11-1-0347. R.G.E. thanks Conicyt Pai/Concurso Nacional de Apoyo al Retorno de Investigadores/as desde el Extranjero Folio 821320024. J.M.F. thanks the support of FAPEMIG.

\end{document}